\documentstyle[eqsecnum,floats,prl,aps,epsf]{revtex}
\begin{document}
\draft
\twocolumn[\hsize\textwidth\columnwidth\hsize\csname
           @twocolumnfalse\endcsname

\title{Gravitational-wave bursts from soft gamma-ray repeaters: Can they be detected?}

\author{H. J. Mosquera Cuesta$^1$, J. C. N. de Ara\'ujo$^1$, 
O. D. Aguiar$^1$, and  J. E. Horvath$^2$}

\address{$^1$Divis\~ao de Astrof{\'\i}sica, Instituto Nacional 
de Pesquisas Espaciais \\ Av. Astronautas 1758,
S\~ao Jos\'e dos Campos, SP, 12227-010, Brazil \\ $^2$Instituto 
Astron\^omico e Geof{\'\i}sico, Universidade de S\~ao Paulo 
\\ Av. Miguel Stefano 4200 - S\~ao Paulo SP, 04301-904, Brazil}


\maketitle

\begin{abstract}
\widetext
 In this letter we suggest a scenario for simultaneous emission of gravitational-wave and $\gamma$-ray bursts (GRBs) from soft gamma-ray repeaters (SGRs). we argue that both of the radiations can be generated by a super-Eddington accreting neutron stars in X-ray binaries. In this model a supercritical accretion transient takes back onto the remnant star the disk leftover by the hydrodynamic instability phase of a low magnetized, rapidly rotating neutron star in a X-ray binary system. We estimate the rise timescale $\Delta t_c = 0.21 \; ms$, minimum mass accretion rate needed to trigger the $\gamma$-ray emission, $\dot{M}_\lambda = 4.5 \times 10^{28} \; g$, and its effective associated temperature $T_{eff} = 740 \; keV$, and the timescale for repeating a burst of $\gamma$-rays $\Delta \tau_R = 11.3 \; yr$. Altogether, we find the associated GW amplitude and frequency to be $h_c = 2.7 \times 10^{-23}/{(Hz)}^{1/2}$ and $f_{gw} = 966 \; Hz$, for a source distance $\sim 55 \; kpc$. Detectability of the pulses by the forthcoming GW anntenas is discussed and found likely.
\end{abstract}

\vskip 2pc
\narrowtext
]
\def\be{\begin{equation}}
\def\ee{\end{equation}}

With the advent of the gravitational-wave astronomy a new window in physics 
is about to be opened. The first GW signal, whenever detected with the new 
generation of observatories such as LIGO, VIRGO, GEO600, TAMA \cite{thorne96},
and/or the fourth generation of resonant-mass detectors like TIGAs 
\cite{Merk-John96,frossati,aguiar,coccia,Harry}, will mark a breakthrough in 
modern astrophysics \cite{thorne95,schutz95}. Most theoretical studies to 
address the prospects have been concentrated in sources like 
coalescing NS binaries \cite{will,damour} and rapidly rotating NSs 
\cite{wagoner,centrella,lai,de-araujo}. Statistical estimates of event rates, 
for both sources, point to an occurrence rate $\sim 3/yr$, in the best cases, 
for distances up to the Virgo cluster \cite{finn93,phinney91,lipunov1}. 
On the other hand, neutron star hydrodynamical instabilities, like 
that one studied by Houser, Centrella \& Smith (1994) (HCS'94) and Lai \& 
Shapiro (1994)\cite{centrella,lai}, should occur quite  
frequently in nature and we focus on the post-hydrodynamical 
instability phase in such systems, with the perspective of modeling SGRs, which may also be interesting GW sources \cite{herman}. 

Because of the physics underlying the exact nature of soft gamma-ray
repeaters (SGRs) is still an unsolved puzzle in high energy astrophysics, the present study is a novel ingredient in our search for understanding these objects. Together with, it is the realization that the foreseeable detection of gravitational waves (GWs), as discussed above, may help underpinning various open problems in both lines of research, yet. The possible confluence of these widely different trends in modern astronomy is an interesting perspective. Motivated by such a worthy possiblity, we propose that a neutron star accretes matter from a companion in a X-ray binary system until becoming a millisecond spinning star. Then, centrifugal effects drive equatorial mass ejection, and form a thick disk surrounding the NS. Various instabilities in this dense structure forces it to fragment into clumps that are later on capture back onto the star. Our model allows us to estimate basic important features of SGRs as its {\it risetime}, minimum mass accretion rate needed to trigger the $\gamma$-ray emission and its effective associated temperature. The time interval to becoming a repeating $\gamma$-ray burster is a worthy by-product of our scenario. as shown below our results are in quite well agreement with observations.

Since most of SGRs observed fluxes indicate a luminosity $\sim 10^{41-42} \; erg/s$ (or probably higher, $\sim 10^{44} \; erg/s$ \cite{FKL95}); well above the Eddington luminosity, it is natural to consider these binary systems in a supercritical accretion regime. In our SGRs model we envision a disk-like structure as being a donut-type thick encircling disc with characteristic dimensions, (see Table-1): $\Delta w = 2 \times R_{ns}$ as the disk width; the difference between the external radius $R_d$; assumed to be that one derived for the final structure in HCS'94 simulations, $R_d \sim 5 R_{ns}$, and 
the internal radius, supposed to be the tidal radius, $R_{tide} \sim 3 R_{ns}$. 
We also assume that the disk density is near that one for the HCS'94 remnant 
disk-like structure, i. e., $\rho_d \sim 5.7 \times 10^{12} \; g/cm^3$ 
\cite{centrella}. For parameters of our NS model see Table-2 \cite{herman}.$\ddag$ 

Fluid structures, like this accretion disk-type object, may undergo various 
instabilities, depending on the dynamical conditions ({\it Jeans, 
Rayleigh-Taylor instability, viscous}). Provided the system is driven by Jean's 
instability, clumps of matter around the central compact body are 
produced. The lengthscale of these 
clumps follows from the Jeans' wavenumber

\be
K{_j^2} ={ {4 \pi G \rho_d} \over {a{_d^2}} }
\ee

where $K{_j}$ is the wavenumber of the perturbation in the matter, $a{_d} = 
{({\Delta p \over \rho_d})}^{1\over 2}$ is the sound speed in the structure and 
$\rho_d$ its mass density. Hence, assuming that 
$a_d \sim  10^{-1} (c) \; cm/s$, we 
find $K_j \simeq 7.3 \times 10^{-7} \; cm^{-1}$, and therefore 
determine an associated length, 
$\lambda_j \sim 8.7 \times 10^{6} \; cm $. 
We attribute the production of a burst to the fall of a set of clumps 
back onto the NS (see below).

In this scenario the rise timescale and the 
time-delay between peaks in the soft 
$\gamma$-ray bursts spectra are determined 
as follows: when fragments fall back onto the NS surface, 
with balistic trajectories, each blob is stretched 
to a needle shape \cite{Ramana}. The time interval for this collision (blob-NS) 
to occur would be

\be
\Delta t_c \approx {\left({ {\lambda/2} \over {(G M_{ns}/ R{^2_{ns}}) } 
}\right)}^{1\over 2}, 
\ee

here $M_{ns}$ and $R_{ns}$ are the mass and radius of the NS, 
respectively. Therefore, we argue this time interval, $\Delta t_c \sim 2.1 
\times 10^{-4} \; s$, is that one characteristic of risetimes in SGRs. For 
comparison, it is worth noticing that the 790305 b 
event had a rise time $\le 0.25\; ms$ \cite{FKL95,Ramana}. The 
characteristic time between intervening peaks in a given burst structure, (the 
timebreak), could roughly be inferred from ${{\tau_s}^{-1}} = K_{j} \times 
V_{orb}$ and turns out to be  $\tau_s \sim 168.2 \; \mu s$. 
Here $V_{orb} = {({ {2 G M } \over { R_\times} } 
)}^{1\over 2} \sim$ (0.273 c) $cm/s$, 
is the orbital velocity of each clump before 
being accreted. This value is over the GRO temporal resolution: $\sim 100 \;\mu s$ \cite{Ramana,FKL95}, and therefore may be observed. In clumped structures such as these 
strong turbulence effects should play an 
important role in driving the disk hydrodynamics. We may 
take these effects into 
consideration in our present simplified description throughout the parameter 
$\beta_{acc}$, which tells us about the 
efficiency of viscosity in governing the structure. 


\begin{table}
\caption[]{\label{tbl-1}}{Parameters of our disc-like structure: mass $M_d$, density $\rho_d$, width $\Delta\omega$, external radius $R_d$, tidal radius $R_{tide}$, orbiting radius $R_{orb}$, disc height-scale $\Delta H$ and Afv\'en radius $R_{acc}$.} 
\vspace{5pt} 
\begin{tabular}{cc}
{Parameter} & {Value} \\
\tableline
$M_d \; [M_\odot]$ & $4.0 \times 10^{-2}$ \\
$\rho_d \; [g/cm^3]$ & $5.7 \times 10^{12}$ \\
$\Delta\omega \; [cm] $ & $2.82 \times 10^6$ \\
$R_d \; [cm]$ & $7.05 \times 10^{6}$ \\
$R_{tide} \; [cm]$ & $4.21 \times 10^{6}$ \\
$R_{orb} \; [cm]$ & $5.64 \times 10^{6}$ \\
$\Delta H \; [cm]$ & $5.0 \times 10^4$ \\
$R_{acc} \; [R_\odot]$ & $1.0 \times 10^3 $ \\
\end{tabular}
\end{table}

After losing enough energy and part of its angular momentum through those 
mechanisms, the clumped disc-like mass distribution falls onto the NS on a 
free-fall timescale. When the blob hits on the NS surface it triggers a strong 
thermal-like burst of $\gamma$-rays. The 
characteristic temperature of the burst can be estimated from 
equating the luminosity of accretion to the radiation power. This yields the 
equation

\be
T_{peak} = {\left( {G \over {4 \pi \sigma_{SB} } } { {\beta_{acc} M_\lambda} 
\over {\tau_{ff}} } {M_{remn} \over { R{^3_{remn}}} } \right)}^{1\over 4}.
\ee

$M_\lambda \approx {1 \over 6} \pi \rho_d ({\lambda} \Delta w \Delta H)$, with 
$\rho_d$ being the disc density; $T_{peak}$ is the radiation temperature, and 
$M_{remn}$ and $R_{remn}$ represent the mass and  radius of the remnant compact 
star. We define $\dot{M_\lambda} \approx { {M_\lambda} \over {\tau_{ff}} }$, 
being ${\tau_{ff}} = {(G \rho)}^{-1/2} \simeq 548 ms$. The equivalent 
accretion rate per year yields 
$\dot{M_\lambda} \simeq {1.1 \times 10^5} \;{M_\odot / yr}$. Assuming that the 
hard component of GRB 790305 b is that one characteristic of all 
SGRs, it can be seen that most of SGRs 
observed thusfar would have had associated mass accretion rates, 
$\dot M_\lambda \sim 4.5 \times 10^{28} \; g/s$. 
We shall assume a value of the parameter 
$\beta_{acc} \sim 0.70$ for the case of accretion onto NSs.$\dag$ 


\begin{table}[htb]
\caption[]{\label{tbl-2}}{Our NS model parameters: period $P$, mass $M_{ns}$, radius $R_{ns}$, angular momentum $J$, rotational to kinetic energy ratio $\beta$. For {\it initial} we mean at the advent of the instability, and for {\it final} the stable stage after it. } 
\vspace{5pt} 
\begin{tabular}{ccc}
{ } & {initial} & {final} \\
\tableline
$P \; [ms]$ & 0.900 & 0.98 \\
$M_{ns} \; [M_\odot]$ & 1.47 & 1.42 \\
$R_{ns} \; [cm]$ & $1.37 \times 10^{6}$ & $1.41 \times 10^{6}$ \\
$J \; [g cm^2/s]$ & $9.84 \times 10^{49}$ & $8.78 \times 10^{49}$\\
$\beta$ & 0.2738 & 0.27 \\
\end{tabular}
\end{table}

By using parameters derived from our scenario (see Table 1), we obtain for 
the burst peak temperature a value $T_{peak} \sim 2.6 \times 10^{10}\; K$, or 
equivalently, $T_{peak} \sim 2.2 \; MeV$. 

Observations of SGRs \cite{Mazets79} have shown that in 
GRB 790305 b there was a 
soft spectral component which contained most of the energy and 90\% of the 
photons between 30 and 2000 $keV$. Nevertheless, for blackbody models of the 
emission processes the effective temperature, i. e., 
that one is measured by the detectors, $T_{eff} 
\approx T_{peak}/3$. Thus, we infer from our model the value  740 $keV$ which is comparable to the hard component observed in SGR 790305 b, $520 \pm 100\; keV$
\cite{Ramana}. This leads us to conclude that the scales of 
temperatures predicted and observed agree with each other. 
Higher accretion rate would yield in stronger 
$\gamma$-ray bursts. That it would be the case, for example, for the expected 
accretion rate, $\sim 8.9 \times 10^{30}\; g/s$, in the HCS'94 ejected 
disk-like matter; we expect the 
maximum of the emission in that situation to peak at $\sim 3.8 \;MeV$ and the 
associated GW pulse in that event to be slightly stronger than 
the one computed for SGRs in the present estimate (see below).

Viscosity torques remove angular momentum forcing the disk eventually to 
fall down 
onto the star after a time $\Delta t_{frag} \sim R{^2_d} / \nu \geq 10^6\; s$, 
from the instant of its formation. The compact body recovers angular momentum, 
and rotates more quickly. The amount of angular momentum restored, 
after the accretion of all the 
clumps of matter, reads

\be
\beta_{acc} \times N \times M_\lambda \times V_{orb} \times R_{orb} \simeq \Delta J_{rec} 
\ee

where $N$ represents the number of clumps, $ \sim K_j \times 2 \pi R_{orb} \sim 
5$, $R_{orb}$ defines the radius for bound orbiting, $\sim 4.0 R_{ns}$, 
(HCS'94), and $\Delta J_{rec}$ the recovered angular momentum. With 
our assumed parameters for the accreted material 
we get $\Delta J_{rec} \sim 5.2 \times 10^{48} \;g cm^2 s^{-1}$. 

The timescale to repeat a $\gamma$-ray burst can be obtained from the amount of 
the angular momentum to be accreted from the main accretion disk (the donor 
star) in order to reentering into a new unstable phase. To estimate such 
contribution we first calculate the value of the 
energy parameter, $\beta$, after the transient. This yields

\be
\beta = f \frac {J{^2_{stable}/2 I_{stable}}}{G M{^2_{stable}}/R_{stable}} = 
0.27,
\ee

where we have introduced the factor $f$ to measure 
the departure of the star from the dynamical instability, which lies 
the interval 
$0.3 \le f \le 0.6$, and parametrizes the actual star's non-sphericity and 
polytropic index of its equation of state 
in the post-hydrodynamic instability phase. Here the 
subscript word {\it stable} 
stands for secular phase. The lower limit for $f$ indicates the star has 
encountered the {\it secular instability}. The higher one guarantees the 
achieving of the {\it dynamical limit}. 
In the present situation we expect this factor $f$ to be $\sim 0.6$. The 
$\beta$ value for the stable phase, in eq.(5), determines the amount of angular 
momentum to be gained from the primary star at the characteristic rate of 
accretion in such systems $\dot M \sim  (10^{-7}-10^{-5}) \;M_\odot / yr$ \cite{nomoto}.  

The next step is to determine $\Delta \tau_{R}$, the time to repeat a burst. 
Within a burster phase, we use the relation between the 
inward angular momentum, 
$\dot J^+$, and the amount of 
angular momentum to be injected from the main accretion disk, 
$\Delta J_{R}$ (so that the star 
becomes unstable again at $\beta \sim 0.2738$), to write

\be
\dot J^+ \times \Delta \tau_{R} \simeq \beta_{acc} \times \Delta J_{R},
\ee

where $\Delta \tau_{R}$ defines the timescale for bursting again, and $\dot 
J^+ \equiv \dot{M_d} {(G M_{stable}R_{acc})}^{1/2}$ \cite{Shap-Teuk}. Here 
$\dot{M}_{d}$ defines the rate of mass accretion from the main disk, $R_{acc}$ corresponds to the distance the accreted matter 
will come from, the Afv\'en radius for the low $\vec B$ NS in the system, $\sim 10^{3} {R_\odot}$. Thus, if we 
impose for our model NS an accretion rate $\dot M_d \sim 10^{-6}\; {M_\odot}/yr$ we 
find $\Delta \tau_{R} \sim  11.3 \;yr$. This result is in the observed 
ballpark figure of SGRs active phase timescales 
$\sim $ 10 years. Particularly, the source SGR 
1900+14 was observed for the first time in 1979 \cite{Mazets79}. The SGR 1806-20 was observed in 1987 \cite{Ramana}. Both became active again in the nineties as confirmed by observations from BATSE (CGRO) and ASCA 
\cite{Kouve93,Kouve94,FKL95}. SGR 0526-66 was detected in December 1981 and April 
1983, according to observations from the spacecrafts Venera 13 and 14 
\cite{Ramana}. In this picture, the accretion rate must  become strongly 
supercritical in the last source, 
$\dot M \sim 10^{-4}\; {M_\odot}/yr$, so that the time scale would turn out to be 
$\sim 0.11\; yr$, close to observed time intervals of gamma-ray 
recurrences in some SGRs, which are $\sim$ 39 days \cite{Ramana}. 
This result is along the lines of King 
et al. (1997) model \cite{King97}, and suggests that it can find SGRs systems 
undergoing episodes of irradiation-driven supercritical mass transfer (IDSMT) 
during short time spans, depending on binary orbital parameters, with 
recurrences occurring at time steps as short as $\sim 40$ days. 

Now, for the case of the theoretical HCS'94 NS remnant, and using the same 
distance and rate 
of accretion as above, with a stable mass for the NS $\sim 1.38 {M_\odot}$, we 
obtain for $\Delta \tau_{R} \sim 2.4 \times 10^3\; yr$. 
Hence, if the accretion 
process were supercritical, as in systems undergoing IDSMT, 
the resulting timescale for 
repeating a burst would be $\sim  24 \;yr$, what is nearly two orders of magnitude longer than that one for {\it Soft Gamma-Ray Repeater Low Mass X-Ray Binaries} 
(SGR-LMXBs), just estimated in the last paragraph.

Finally, we wish to determine the characteristic GW amplitude associated with 
the burst of $\gamma$-rays. In our model, it is the sudden accretion of matter 
what triggers the emission of a pulse of GWs during the transition phase to a 
stable rotating NS, 
through the fluid modes {\it f}. To estimate the GW burst amplitude we can 
compare the GW flux received at Earth to the total gravitational luminosity 
radiated away by the source \cite{schutz95}, extracted from the gravitational potential energy 
of the orbiting disk-like
structure

\be
\frac {c^3}{16 \pi G} {|\dot h|}^2 = \frac {1}{4 \pi D^2} \frac {\Delta 
E_{GW}}{\Delta t}
\ee

where $\Delta E_{GW}$ represents the energy flowing into GWs, and is defined as 
$\Delta E_{GW} = \Delta L_{GW} \times \Delta t_{c}$, with $\Delta t_c$ 
defined in eq.(2). The gravitational luminosity $\Delta L_{GW}$ is given by 
\cite{LPPT}

\be
\Delta L_{GW} = K \times {\frac {G {M_\lambda}^2 L^4} {5 c^5 {(\Delta 
t_{c})}^6 } }.
\ee

where K is a constant:  
$10^{-2} \le K \le 10$ \cite{ruffini94}. 
We assume the value $K \, = \, 0.5$, the mass being 
$M_\lambda = 3.6 \times 10^{30}\; g$; the one that 
triggers the emission of the highest energy photons, 
and the parameter $L$ is the radius
of the orbit of that mass around the remnant, 
$\sim 4 R_{ns}$. Introducing these 
figures in the equation

\be
h_c = {\left[ {\frac {4 G}{c^3 D^2} } {\left( \Delta L_{GW} \times {(\Delta 
t_c)}^2 \right)} \right]}^{1/2},
\ee

we get for the GW amplitude the value  $h_c \sim 8.25 \times 10^{-22}$, for a 
source distance, $D$, of 55 $kpc$.

In order to be detected by the TIGAs network, the SGRs characteristic 
GW amplitude, when normalized to its frequency Fourier transform 
(the square root of the 
characteristic spectral frequency distribution), our estimated GW-amplitude 
reduces to $h_c = 2.7 \times 10^{-23}/{(Hz)}^{1/2} $. 
We propose that the GW-frequency should be 
correlated with the timescale during which half of the infalling clumps of 
matter impact the NS surface, i. e., $f_{gw} \sim 1/ {(5 \times \Delta t_c)}$ = 
966 $Hz$. This timescale specifies the elapsed time until 
a "clean" GW signal from the system is 
settled down in the GW detector. The frequency sensitivity bandwidth being 
$\Delta f_{gw} = 2 \mu f_{gw} \sim 30 \;Hz$, with $\mu $ the inverse of the {\it 
two-mode} transducer mass ratio. 
Such frequency bandwidth is expected to be achieved for most of the 
planned resonant-mass detectors \cite{Merk-John96,Harry}. The 
frequency-averaged 
GW amplitude is certainly under the quoted TIGA's sensitivity cut off (see 
Figure 1). 

As a check of the calculation just done, we can compute the GW characteristic 
amplitude with the Wagoner's (1984) equation \cite{wagoner} rescaled for the 
actual accretion mass rate, ${\dot{M_\lambda}}$, in our scenario 

\be
h = \sqrt{5} \times 10^{-26} {\left( { {1{kpc}} \over D} \right)} {\left( 
{{kHz} \over {f_{gw} m}} \right)}^{1\over 2} {\left( { {\dot{M_\lambda}} 
\over {10^{-8} \, {M_\odot/year}} } \right)}^{1\over 2}.
\ee

This yields $h_c = 3.0 \times 10^{-23}/{(Hz)}^{1/2} $. The calculation 
for the whole range of frequencies (1-3000) Hz are presented in Figure-1. 
All of the plotted values were normalized for the 
frequency spectral distribution 
(Hz)$^{-1/2}$ of the outgoing GW burst, in order to be compared to 
the TIGAs network and LIGO sensitivities, 
as shown in Figure-1.


\begin{figure}[htb]
\centerline{\vbox{\epsfxsize=8cm\epsfbox{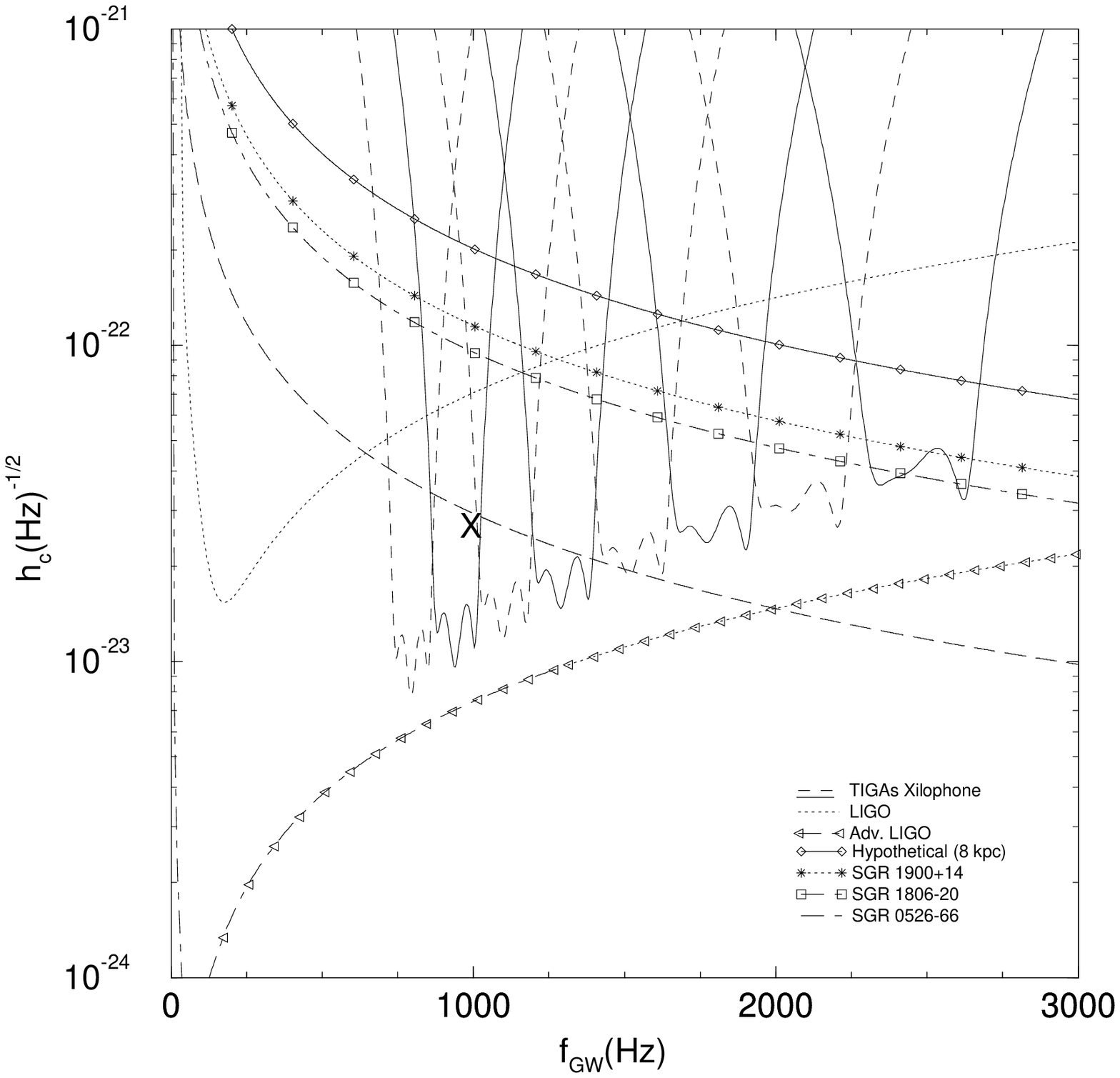}}}
\caption[]{\label{fig-2}}{Locus of the dimensionless GW characteristic peak 
amplitude, $h_c$ (log), plotted against the wave frequency, f$_{GW}$ (linear), 
for the three up-to-date known soft $\gamma$-ray repeaters ({\it 
distance$_{[kpc]}$,$\gamma$-flux}$_{[erg/(cm^2 s)]}$): 
SGR 0526-66 (55,$ \sim 1.5 \times 10^{-3}$), SGR 1900+14 
(14,$ \sim  4.0 \times 10^{-4})$ and SGR 1806-20 (17,$\sim 2\times 10^{-4})$. 
Also data are plotted for a hypothetical strong $\gamma$-ray source, 
$\gamma$-flux $\sim 5 \times 10^{-2}$ (top-line), placed at 8 $kpc$. The symbol {\bf X} represents the GW characteristics computed with eq.(9).}
\end{figure}


Our model is based on the hypothesis that there exists a peculiar 
class of LMXBs 
where higher accretion rates are at work. The burst duration in 
SGR-LMXBs should 
be associated with the total accreted mass for the transient. As shown above, 
the time span for re-bursting can be inferred to from the burst 
temperature. During quiescent 
stages, assuming an Eddington's accretion rate and a source distance $\sim 55 \; kpc$, their X-ray fluxes must be around $3.4 \times 10^{-16}\; erg/(cm^2 s)$. 
These fluxes are marginally detectable by Beppo-SAX or ASCA X-ray detectors. For galactic distances, $\sim 10 \;kpc$, those sources would certainly be detectable by such capabilities. For comparison, observed values for the SGRs X-ray luminosities are $L_x = 7 \times 10^{35} \; erg/s$ and $ L_x = 3 \times 10^{34} (D/8 kpc) \; erg/s$, for SGR 1900+14 and 1806-20, respectively \cite{duncan}. 

As a point aside, there are left two further pathways for the thick disk to evolve. Firstly, it may not fragment apart anyway. Thus, the whole disk infalls at once onto the NS triggering, therefore, more powerful bursts of both of the radiations. Secondly, had fragments it produces very small lenghtscales, then, instead of being observed as a $\gamma$-rays train of pulses it should appear as persistent bursts of hard or even soft X-rays, with characteristic timescale $\sim 1\; month$, and time recurrences far too short, $\sim 3\; mins$, alike to the ones observed by BATSE from the source GRO J1744-28 \cite{fishman,kouve}. Sources such as these may prove interesting astrophysical labs to test this scenario. Also do numerical simulations of thick dense disk neutron star interactions taking into account relativistic effects, or instead observations of accretion X-ray systems in which accretion rates $\sim 10^{-6} \; M_\odot/yr$ are seemingly common$\ddag$, or even such systems as those simulated by Nomoto (1986) and Saio \& Nomoto (1998) \cite{nomoto,saio}.
 
Our main result is displayed in Figure-1 which shows that the ``xilophone" of 
TIGAs, with a bandwidth $\sim 30 \;Hz$, may observe GWs through the 
NS fluid modes 
from SGRs, whenever they radiate at frequencies lower than 1.4 $kHz$, for 
distances as far as the LMC. As an example of the 
detectability of these signals by ressonant-mass 
detectors it is plotted the pulse computed with eq.(7), marked with 
an (X), which corresponds to $h_c = 2.7 \times 10^{-23}/{(Hz)}^{1/2}$ and a $f_{gw} = 966 \; Hz$. Interferometric GW detectors like LIGO, for example, 
may observe such systems provided they emit 
at mid frequencies, $\leq$ 640 $Hz$ (first LIGO), and up 
to 2000 $Hz$ (advanced LIGO), for the same distances. Given the potential implication for the physics of both  
SGRs and GWs, detailed studies of the model would be needed.

\bigskip



{$\dag$ {Note that this choice guarantees us that the 
dynamically unstable NS gets rid of sufficient angular momentum so as to 
becoming itself {\it secularly} unstable (see also Shapiro \& Teukolsky 1983).}}

{$\ddag$ {More details on tables 1 and 2, and also on suggested modelling and  observational work will be given in a work in preparation.}}



\begin{references}

\bibitem{thorne96}
K. S. Thorne, in Proceedings of the {\it Chandrasekhar Symposium}, ed. R. Wald, 
in press (1996).

\bibitem{Merk-John96} 
S. Merkowitz and W. W. Johnson, Phys. Rev. D {\bf 53}, 10, 5377 (1996). 

\bibitem{frossati} 
G. Frossati and E. Coccia, Cryogenics, {\it ICEC} 15,  Supplement, {\bf 34}, 9 (1994), 
J. Low. Temp. Phys., 101, 81 (1995).

\bibitem{aguiar} 
O. D. Aguiar, et al., in{\it  Abst. 13th Int. Conf. on  General Relativity and Gravitation}, eds. Lamberti P. W., Ortiz O. E., FAMAF C\'ordoba, Argentina, p. 455 (1992).

\bibitem{coccia} 
E. Coccia and V. Fafone, in Proceedings of the {\it First 
International Workshop on Omni-directional Gravitational-wave Detectors}, S\~ao 
Jos\'e dos Campos, Brazil, May 26-30, Eds. Velloso, W., Magalh\~aes, N. and 
Aguiar, O. D. (1996).

\bibitem{Harry}
G. Harry, Th. Stevenson and H. J. Paik, PRD {\bf 54}, 4, 2409 (1996).

\bibitem{thorne95} 
K. S. Thorne, in Proceedings of IAU Symposium {\bf 165}, {\it Compact Stars in Binaries}, eds. J. Van Paradijs, E. Van del Heuvel, and E. Kuulkers, Kluwer Academic Publishers (1995).

\bibitem{schutz95} 
 B. F. Schutz, in Proceedings of the {\it Les Houches School on Astrophysical 
Sources of Gravitational Radiation}, Ec\'ole de Physique Des Houches, Eds. J-P. Lasota and J-A. Marck, C. U. P. (1996).

\bibitem{will}
C. Will, in {\it Relativistic Cosmology}, ed. M. Sasaki, Universal 
Academy Press (1994).

\bibitem{damour}
L. Blanchet, T. Damour, and G. Schafer, MNRAS, {\bf 242}, 289 (1990). 

\bibitem{wagoner}
R. V. Wagoner,  ApJ, {\bf 278}, 345 (1984).

\bibitem{centrella} 
J. L. Houser, J. M. Centrella and S. C. Smith, Phys. Rev. Letts. {\bf 72}, 9, 1314 
(1994).

\bibitem{herman}
H. J. Mosquera Cuesta, J. C. Neves de Ara\'ujo, O. D. Aguiar and J. E. Horvath, in preparation (1998 b).

\bibitem{lai} 
D. Lai, and S. L. Shapiro, ApJ {\bf 437}, 742, 1994, ApJ {\bf 420}, 811-829 (1994), ApJ 
{\bf 442}, 259 (1995).

\bibitem{de-araujo} 
J. C. N. de Ara\'ujo, et al., MNRAS,  {\bf 271}, L31-L33 (1994).

\bibitem{finn93}
l. S. Finn, . and  D. F. Chernoff, Phys. Rev. D {\bf 47}, 2198 (1993).

\bibitem{phinney91}
 E.S. Phinney, ApJ {\bf 380}, L17 (1991).

\bibitem{lipunov1} 
V. M. Lipunov, K. A. Postnov, and M. E. Prokhorov, H. E. Jorgensen, ApJ {\bf 453}, 593 
(1995).

\bibitem{FKL95} 
G. L. Fenimore, R. Klebesadel and U. Laros, ApJ {\bf 460}, 964 (1996).

\bibitem{Ramana} 
P. V. Ramana-Murthy and A. Wolfendale, {\it Gamma-Ray Astronomy}, C. U. P., Revised Edition (1993).

\bibitem{Shap-Teuk}
S. S. Shapiro, and S. Teukolsky, {\it Black Holes, White Dwarfs and Neutron 
Stars, The physics of Compact Objects}, Wiley, New York (1983).

\bibitem{Mazets79} 
E. P. Mazets, S. V. Golenetskii, V. N. Ilinskii, R. L. Aptekar and Yu. A. 
Guryan, Nature, {\bf 282}, 587 (1979).

\bibitem{Kouve93} 
C. Kouveliotou, et al., Nature, {\bf 362}, 728 (1993).

\bibitem{Kouve94} 
C. Kouveliotou, et al., Nature, {\bf 368}
, 125 (1994).

\bibitem{King97} 
A. R. King, ApJ. {\bf 482}, 919-928, (1997).

\bibitem{LPPT}
Lightman, R, Price, R., Press, W. and Teukolsky, S., {\it Problem Book in 
Relativity} (1973).

\bibitem{ruffini94} 
H. C. Ohanian ans R. Ruffini, {\it Gravitation and Spacetime}, $2^{nd}$ edition, Norton (1994).

\bibitem{duncan}
C. Thompson and R. C. Duncan, ApJ, {\bf 473}, 322 (1996).

\bibitem{fishman}
J. G. Fishman,  et al., {IAU}, Circular {\bf 6272} (1995)

\bibitem{kouve}
C. Kouveliotou, et al., {Nature}, {\bf 379}, 799 (1996)

\bibitem{nomoto}
K. Nomoto, {\it Collapse of Accreting Carbon-Oxigen White Dwarfs Induced by Carbon Deflagration at High Density}, in proceedings of {\bf Accretion Processes in Astrophysics}, Eds. J. Audouze \& J. Tran Thanh Van, Editions Fronti\`eres, (1986). 

\bibitem{saio}
H. Saio \& K. Nomoto, {\it Inward Propagation of Nuclear-Burning Shells in Merging C-O and He White Dwarfs}, astro-ph/9801084, (1998).

\end{references}
\end{document}